\documentclass[12pt,superscriptaddress,nofootinbib]{revtex4-2} 
\usepackage{ragged2e}

\usepackage{multirow}

\usepackage{amsmath}  % needed for \tfrac, \bmatrix, etc.
\usepackage{mathtools}
\usepackage{amsfonts} % needed for bold Greek, Fraktur, and blackboard bold
\usepackage{graphicx} % needed for figures
\usepackage{xcolor}

\usepackage{array}
\usepackage{makecell}

\usepackage{footnote}

\usepackage[justification=justified,labelfont=bf,format=plain]{caption}
\DeclareCaptionFormat{myformat}{\fontsize{10}{10} \selectfont#1#2#3}
\usepackage{hyperref}

\begin{document}

\title{On the Derivation of \\ the Cosmological Gurzadyan's Theorem}

\author{Trung V. Phan}
\email{{tphan@natsci.claremont.edu}}
\affiliation{Department of Natural Sciences, Scripps and Pitzer Colleges, \\ Claremont Colleges Consortium, Claremont, CA 91711, USA}

\date{\today}

\maketitle

In cosmology, the Gurzadyan's theorem \cite{gurzadyan1985cosmological} identifies the most general force law consistent with the finding of Newton’s first shell theorem \cite{newton1833philosophiae} -- that a spherical symmetric mass exerts the same gravitational force as a point mass at its center. This theorem has found important applications in cosmological modeling \cite{vedenyapin2021generalized,chardin2021mond}, particularly in the context of MoND (Modified Newtonian Dynamics) \cite{milgrom1983modification}, which has recently gained renewed attention as a potential alternative to dark matter. The derivation given by Gurzadyan \cite{gurzadyan1985cosmological} is written in an extremely concise and dense style typical of theoretical work from the former USSR, making it difficult to follow. Recent proofs of the theorem based on power-series methods \cite{reed2022note,Carimalo2023} offer valuable perspectives, though they differ from the original derivation, which is based on \textit{perturbation analysis}. Our note aims to clarify the underlying logic in a pedagogical way -- accessible to advanced high school or undergraduate students -- while preserving conceptual clarity and mathematical elegance of the his insight.

\begin{figure}[!htbp]
\includegraphics[width=\textwidth]{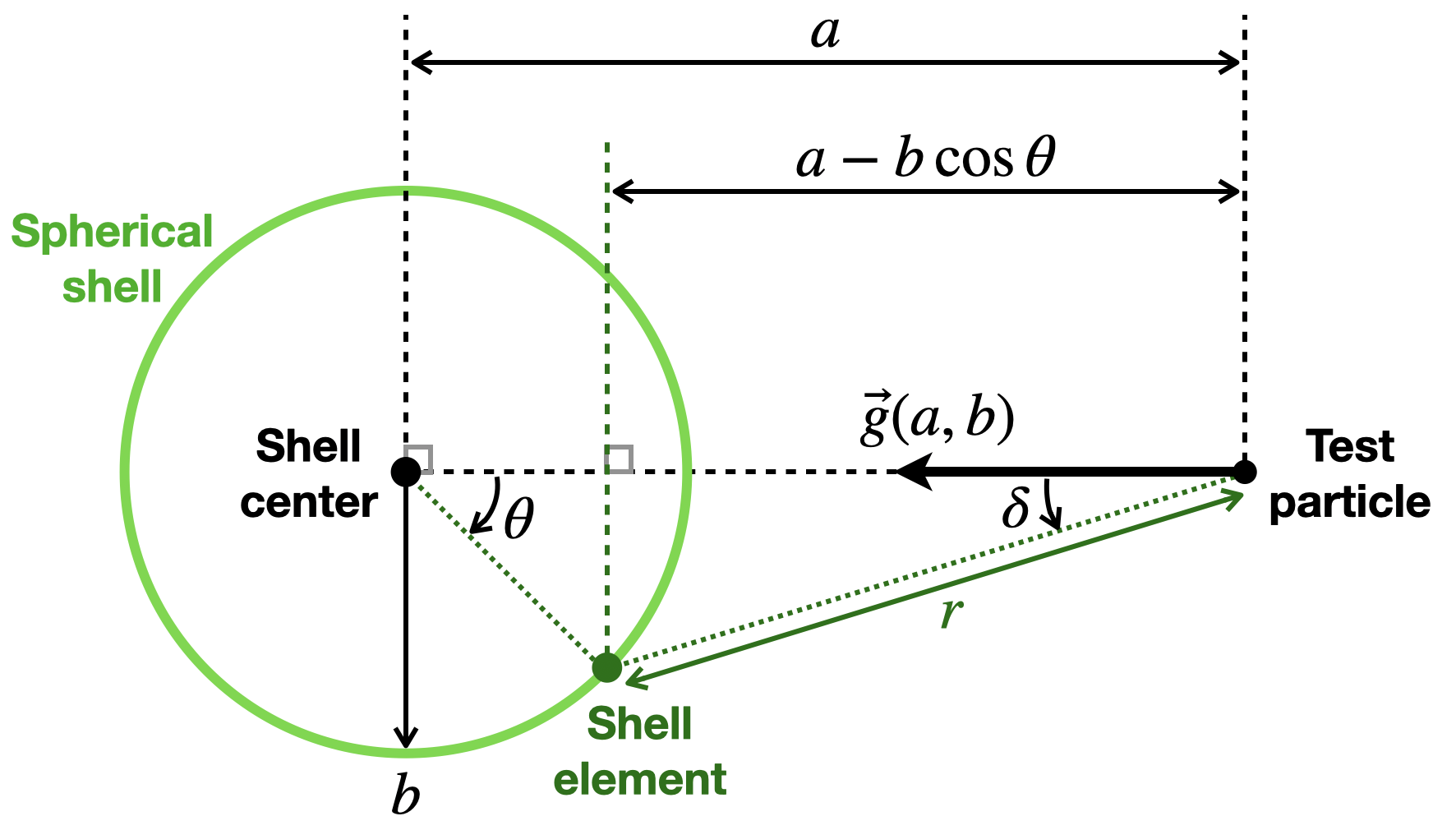}%
\caption{Schematic of the spherical shell and test particle configuration used to calculate the gravitational acceleration $\mathrm{g}(a,b)$ in Eq. \eqref{shell}.}
\label{fig001}
\end{figure}

To find the force law in the Gurzadyan's theorem, it suffices to consider a spherical shell instead of the full volume. This follows from the additive nature of the interaction, which allows the total force to be built from contributions of individual infinitesimal-thin shells. If the theorem holds for a single uniform shell -- producing the correct effect on a test particle outside -- it must also hold for any spherically symmetric distribution. 

Consider a spherical shell of radius $b$, centered a distance $a>b$ from a test particle (see Fig. \ref{fig001}A), and having total unit mass -- the mass density per unit surface area is thus $\mu = 1/4\pi b^2$. We use spherical coordinates $(\rho,\theta,\phi)$ with the origin located at the center of the shell, where $\rho$ is the radial distance, $\theta=[0,\pi]$ is the polar angle measured from the axis connecting the shell center to the test particle, and $\phi=[0,2\pi]$ is the azimuthal angle. Let $r$ be the distance between the test particle and an infinitesimal mass element on the shell, each element exerts a force along $\vec{r}$. Due to symmetry, the transverse components cancel upon integration, leaving only radial $\rho$-components along the axis connecting the shell center and the test particle. The total gravitational acceleration at the position of the test mass, denoted $\mathrm{g}(a,b)$, depends on the distance $a$ and the shell radius $b$, and is given by:
\begin{equation}
\begin{split}
    \mathrm{g}(a,b) &= \mu \int^{\pi}_0 b d\theta \int^{2\pi}_0 b \sin\theta d\phi \ F(r) \cos\delta 
    \\
    &= \frac{1}2 \int^{\pi}_0 d\theta \sin\theta \ F(r) \cos\delta \ ,
\end{split}
\label{shell}
\end{equation}
where $r$ and $\delta$ are calculated from:
    $$r = \left( a^2 + b^2 - 2 ab \cos\theta \right)^{1/2} \ , \  \cos\delta = \frac{a-b\cos\theta}{r} \ . $$
Here, the force law is represented by $F(r)$, such that the interaction force between any pair of masses is given by their product multiplied by this function. 

We seek the most general form $F(r)$ that is consistent with the finding of Newton's first shell theorem \cite{newton1833philosophiae}. For our setting, this translates to the condition for the {\color{black}gravitational} acceleration calculated in Eq. \eqref{shell}:
\begin{equation}
    \mathrm{g}(a,b)\Big|_{b<a}=\mathrm{g}(a,0) \ .
\label{newton}
\end{equation}
If this equation holds for all $b<a$, then it must also hold in the limit $b\rightarrow 0$, where $b$ can be treated as a perturbative parameter. We can then {\color{black} Taylor's expand $\mathrm{g}(a,b)$ as:
\begin{equation}
    \mathrm{g}(a,b) \approx \mathrm{g}(a,0) + b \mathrm{g}_1(a) + \frac12 b^2 \mathrm{g}_2(a) + \mathcal{O}(b^3) \ ,
\label{order}
\end{equation}}
and then set all $\mathrm{g}_{n}(a)=0$. Note that, Eq. \eqref{shell} shows that $\mathrm{g}(a,b)$ is an even function of $b$, i.e. satisfying
$\mathrm{g}(a,b)=\mathrm{g}(a,-b)$. Thus, in Eq. \eqref{order}, the odd-order term should always vanish $\mathrm{g}_{n\in 2\mathbb{N}-1}(a)$. Setting $\mathrm{g}_{n\in 2\mathbb{N}}(a)=0$ gives us infinitely many differential equations of $F$. The leading-order contribution in the perturbative expansion is the $2$nd-order term.

We expand the integrand in Eq. \eqref{shell}, i.e. $F(r)\cos\delta$, to the $2$nd-order in $b$:
{\color{black}\begin{equation}
\begin{split}
    & F(r)\cos\delta \approx F(a) + b\cos\theta F'(a)
    \\
    & \ \ \ \ - \frac{b^2}{2a^2} \Big[ \sin^2\theta F(a) - a\sin^2\theta F'(a)  - a^2 \cos^2\theta F''(a) \Big] + \mathcal{O}(b^3) \ ,
\label{Taylor_of_Int}
\end{split}
\end{equation}
in which} the terms $F'$, $F''$ are $1$st- and $2$nd-derivatives of $F$. {\color{black} A more detailed treatment of this step is provided in the Appendix \ref{app:A}.} With this, we do the $\theta$-integration in Eq. \eqref{shell} an then match with the order expansion in Eq. \eqref{order} to obtain:
\begin{equation}
\begin{split}
    \mathrm{g}(a,0)=F(a) \ & , \ \mathrm{g}_1(a)=0 \ ,
\\
    \mathrm{g}_2(a)= \frac1{3a^2} \Big[ 2F(a) & - 2aF'(a) - a^2 F''(a) \Big] \ .
\end{split}
\label{coeffs}
\end{equation}
{\color{black} We outlined the derivation for this step in Appendix \ref{app:B}.} Setting $\mathrm{g}_2(a)=0$ is equivalent to solving Eq. (4) of the original derivation in \cite{gurzadyan1985cosmological}, i.e.
\begin{equation}
    2F(a) - 2aF'(a) - a^2 F''(a) = 0 \ .
\label{gurz}
\end{equation}
It is unfortunate that the steps leading from Eq. \eqref{shell} and Eq. \eqref{newton} to Eq. \eqref{gurz} are completely omitted in \cite{gurzadyan1985cosmological}. However, we have now reconstructed them.

Eq. \eqref{gurz} provides a necessary condition on 
$F$, but not a sufficient one, as 
$F$ must also satisfy an infinite set of additional differential equations of the form $\mathrm{g}_{n\in2\mathbb{N}+2}(a)=0$. However, once a ``candidate'' solution $F$ is obtained by solving Eq. \eqref{gurz}, we can substitute it back into Eq. \eqref{shell} for general 
$b$ and verify whether Eq. \eqref{newton} holds. If it does, then the function $F$ satisfies all required conditions non-perturbatively and is the solution we seek.

The general solution for Eq. \eqref{gurz}, which is a $2$nd-order homogeneous linear differential equation with variable coefficients and can be solved systematically \cite{strang1991calculus}, has the form:
\begin{equation}
    F(r) = Ar + Br^{-2} \ ,
\label{gurz_sol}
\end{equation}
which is the combination between Coulomb's law $F(r)\propto r^{-2}$ and Hook's law $F(r)\propto r$. Plugging Eq. \eqref{gurz_sol} into Eq. \eqref{shell} does indeed result in Eq. \eqref{newton}, which means it is the most general force law consistent with the finding of
Newton’s first shell theorem \cite{newton1833philosophiae}. This is the statement of the cosmological Gurzadyan's {\color{black}theorem} \cite{gurzadyan1985cosmological}.

We would like to point out that Gurzadyan’s theorem had already been studied, proven, and even generalized \cite{sneddon1949camb,kilmister1974newton,chapman1983gravity,barnes1984gazette} prior to the publication of \cite{gurzadyan1985cosmological}. It would be interesting to explore whether similar theorems might arise in exotic spaces -- such as fractal geometries -- where conventional calculus breaks down \cite{tarasov2005continuous,tarasov2005possible,phan2024vanishing} and new forms of long-range interaction may emerge.

Proving the cosmological Gurzadyan's theorem was part of the training for the Vietnamese IPhO (International Physics Olympiad) team in 2024. We thank Tran D. Huy for helpful discussions, and the xPhO club for encouraging us to share this with a broader audience.

\appendix

{\color{black}
\section{Derive Eq. \eqref{Taylor_of_Int} \label{app:A}}

We want to find the Taylor's expansion in $b$ of the integrand $F(r) \cos\delta$ in Eq. \eqref{shell}, so let us begin by finding the series expansions of the central force $F(r)$ and the trigonometric factor $\cos\delta$.

First, we note that the distance $r$ between a shell element and the test particle can be expanded in powers of $b$ using a Taylor series as follows:
\begin{equation}
    r \approx r\Big|_{b=0} + b \partial_b r\Big|_{b=0} + \frac12 b^2 \partial^2_b r\Big|_{b=0}  + \mathcal{O}(b^3) \ .
\label{r_expand}
\end{equation}
Given that $r=(a^2 + b^2 - 2ab\cos\theta)^{1/2}$ as mentioned in Eq. \ref{shell}, $r\Big|_{b=0}=a$, and we can calculate its derivatives to arrive at:
\begin{equation}
\begin{split}
    \partial_b r \Big|_{b=0} &= \frac{-a\cos\theta + b}{(a^2 + b^2 - 2ab\cos\theta)^{1/2}} \Big|_{b=0} = -\cos\theta \ ,
\\
    \partial_b^2 r \Big|_{b=0} &= \frac{a^2\sin^2\theta}{(a^2 + b^2 - 2ab\cos\theta)^{3/2}} \Big|_{b=0} = \frac{\sin^2\theta}{a} \ .
\end{split}
\end{equation}
Therefore, Eq. \eqref{r_expand} can be rewritten as:
\begin{equation}
    r \approx a - b \cos\theta + \frac{b^2}{2a} \sin^2\theta + \mathcal{O}(b^3) \ . 
    \label{r_expand_explicit}
\end{equation}
Then, we Taylor's expand the central force $F(r)$ around $r=a$:
\begin{equation}
    F(r) \approx F(a) + (r-a)F'(a) + \frac12 (r-a)^2 F''(a) + \mathcal{O}\left[ (r-a)^3 \right] \ .
\label{F_expand}
\end{equation}
Applying the result of Eq. \eqref{r_expand_explicit} and organizing by powers of
$b$, we find:
\begin{equation}
    F(r) \approx F(a) - b\cos\theta F'(a) + \frac{b^2}{2a^2}\Big[ a \sin^2\theta F'(a) + a^2\cos^2\theta F''(a) \Big] + \mathcal{O}(b^3) \ .
\end{equation}

Next, we need to expand $\cos\delta$, mentioned in Eq. \eqref{shell}, in powers of $b$. From the expansion of $r$ given in Eq. \eqref{r_expand_explicit}, we can calculate its reciprocal:
\begin{equation}
    \frac1r \approx \frac{1}{a} + \frac{b}{a^2}\cos\theta + \frac{b^2}{a^3}\left(1 - \frac32 \sin^2\theta \right) + \mathcal{O}(b^3) \ ,
\end{equation}
and thus obtain:
\begin{equation}
    \cos\delta = \frac{a-b\cos\theta}{r} \approx 1 - \frac{b^2}{2a^2}\sin^2\theta  + \mathcal{O}(b^3) \ .
\end{equation}
Together with Eq. \eqref{F_expand}, we get:
\begin{equation}
\begin{split}
    &F(r) \cos\delta \approx F(a) - b\cos\theta F'(a) 
    \\
    & \ \ \ \ + \frac{b^2}{2a^2}\Big[ -\sin^2\theta F(a) + a \sin^2\theta F'(a) + a^2\cos^2\theta F''(a) \Big] + \mathcal{O}(b^3) \ ,
\end{split}
\end{equation}
which is identical to Eq. \eqref{Taylor_of_Int}.

\section{Derive Eq. \eqref{coeffs} from Eq. \eqref{Taylor_of_Int} \label{app:B}}

We can do the integration in Eq. \eqref{shell} order by order in powers of $b$, since the integrand is already expressed in  Eq. \eqref{Taylor_of_Int} as a power series in $b$:
\begin{itemize}
    \item For the $b^0$ order:
    \begin{equation}
        \frac12 \int^\pi_0 d\theta \sin\theta \ F(a) = F(a) \ .
    \end{equation}
    \item For the $b^1$ order:
    \begin{equation}
        \frac{b}2 \int^\pi_0 d\theta \sin\theta \ \cos\theta F'(a) = 0 \ .
    \end{equation}
    \item For the $b^1$ order, there are $3$ contributions:
    \begin{equation}
    \begin{split}
        -\frac{b^2}{4a^2} \int^\pi_0 d\theta \sin\theta \ \sin^2\theta F(a) &=-\frac{b^2}{3a^2} F(a)
        \\
        \frac{b^2}{4a} \int^\pi_0 d\theta \sin\theta \ \sin^2\theta F'(a)&= \frac{b^2}{3a} F'(a)
        \\
        \frac{b^2}{4} \int^\pi_0 d\theta \sin\theta \ \cos^2\theta F''(a) &= \frac{b^2}6 F''(a) \ .
    \end{split}
    \end{equation}
\end{itemize}
Adding them all up, we get:
\begin{equation}
    g(a,b) = \frac{1}2 \int^{\pi}_0 d\theta \sin\theta \ F(r) \cos\delta \approx F(a) - \frac{b^2}{6a^2} \Big[ 2F(a) - 2 a F'(a) - a^2 F''(a) \Big] +  \mathcal{O}(b^3) \ .
\end{equation}
This result can then be matched with Eq. \eqref{order} to give Eq. \eqref{coeffs}.

}

\bibliography{main}
\bibliographystyle{apsrev4-2}

\end{document}